\documentclass[apj]{emulateapj}
\submitted{Submitted to The Astrophysical Journal Letters on 2020 June 8.  Accepted on 2020 June 16. }

\usepackage{color}

\shorttitle{Spin-orbit alignment of the \object{$\beta$\,Pictoris} planetary system}
\shortauthors{Kraus et al.}

\begin{document}


\title{Spin-orbit alignment of the $\beta$\,Pictoris planetary system}



\author{
Stefan~Kraus\altaffilmark{1}, 
Jean-Baptiste~Le\,Bouquin\altaffilmark{2},
Alexander~Kreplin\altaffilmark{1},
Claire L.~Davies\altaffilmark{1},
Edward~Hone\altaffilmark{1},\\
John D.~Monnier\altaffilmark{3},
Tyler~Gardner\altaffilmark{3},
Grant~Kennedy\altaffilmark{4},
Sasha~Hinkley\altaffilmark{1}
}

\email{s.kraus@exeter.ac.uk}

\affil{
$^{1}$~School of Physics, Astrophysics Group, University of Exeter, Stocker Road, Exeter EX4 4QL, UK\\
$^{2}$~Universite d'Grenoble Alpes, CNRS, IPAG, 38000 Grenoble, France\\
$^{3}$~Department of Astronomy, University of Michigan, 311 West Hall, 1085 South University Ave, Ann Arbor, MI 48109, USA\\
$^{4}$~Department of Physics, University of Warwick, Coventry, CV4 7AL, UK
}


\begin{abstract}
A crucial diagnostic that can tell us about processes involved in the formation and dynamical evolution 
of planetary systems is the angle between the rotation axis of a star and a planet's orbital angular momentum vector
(``spin-orbit'' alignment or ``obliquity'').
Here we present the first spin-orbit alignment measurement for a wide-separation exoplanetary system,
namely on the directly-imaged planet $\beta$\,Pictoris~b.
We use VLTI/GRAVITY spectro-interferometry with an astrometric accuracy of 1\,$\mu$as (microarcsecond)
in the Br$\gamma$ photospheric absorption line to measure the photocenter displacement 
associated with the stellar rotation.
Taking inclination constraints from astroseismology into account, we constrain the 3-dimensional orientation 
of the stellar spin axis and find that $\beta$\,Pic\,b orbits its host star on a prograde orbit.
The angular momentum vectors of the stellar photosphere, the planet, and the outer debris disk
are well-aligned with mutual inclinations $\leq 3\pm5^{\circ}$, which indicates that 
$\beta$\,Pic\,b formed in a system without significant primordial misalignments.
Our results demonstrate the potential of infrared interferometry to measure the spin-orbit
alignment for wide-separation planetary systems, probing a highly complementary regime to 
the parameter space accessible with the Rossiter-McLaughlin effect.
If the low obliquity is confirmed by measurements on a larger sample of wide-separation planets,
it would lend support to theories that explain the obliquity in Hot Jupiter systems with
dynamical scattering and the Kozai-Lidov mechanism.
\end{abstract}

\keywords{
  stars: rotation --- 
  stars: individual ($\beta$\,Pictoris) --- 
  planets and satellites: individual ($\beta$\,Pictoris~b) ---
  planets and satellites: dynamical evolution and stability --- 
  planetary systems ---
  techniques: interferometric}

\section{Introduction}

Some of the earliest theories about the planet formation process have been 
proposed by Kant \& Laplace in the 18th century and were based primarily on the observation 
that the orbits of the solar system planets are well aligned with each other and with the Sun's spin axis.
In the solar system the three-dimensional obliquity angle $\psi$, i.e.\ the angle between the 
rotation axis of a star and the orbit angular momentum vector of the planets, is $\psi \lesssim 7^{\circ}$ \citep{bec05}.
For transiting extrasolar planetary systems, the \textit{sky-projected} obliquity angle ($\beta$) can be
measured with the Rossiter-McLaughlin (RM) effect \citep{que00}.
For the short-period systems where RM measurements are possible, significant misalignments have been found 
in about a third of the systems \citep{win10}, with values ranging from 0$^{\circ}$ (indicating a perfectly aligned prograde orbit) 
to 180$^{\circ}$ \citep[retrograde orbit, e.g.][]{heb11}.
The sky-projected obliquity angle ($\beta$) can then be related to the true obliquity ($\psi$) distribution,
either by measuring the line-of-sight inclination of the stellar rotation axis $i_{\mathrm{s}}$, 
for instance using astroseismology \citep[e.g.][]{wri11,zwi19}, or by invoking statistical arguments \citep[e.g.][]{mun18}.

Several possible dynamical mechanisms have been proposed to explain the observed spin-orbit misalignments.
In one class of theories, the inclination of the planet's orbit is excited through multi-body gravitational interactions, 
such as planet-planet scattering or interactions with stellar-mass companions \citep[][]{cha08,val14}.
On appeal of this theory is that Kozai-Lidov oscillations \& tidal friction can facilitate both inwards migration
while increasing simultaneously the obliquity of the planet orbit, 
thereby explaining key characteristics of the Hot Jupiter population \citep{koz62,fab07}.

Alternatively, it has been proposed that the stellar rotation axis and orbital plane of the 
planet-forming disk might not be aligned from the start. 
Misalignments between the stellar rotation axis and disk might be induced by
magnetic \citep{lai11} or fluid-dynamical effects \citep{rog13}.
The disk might also be tilted due to turbulent motions in the star-forming cloud \citep{fie15}, 
due to the complex dynamical interactions in the dense clusters where stars form \citep{bat10},
or disk tearing effects induced by the dynamical interactions in multiple systems \citep{nix12,kra20}.

In order to distinguish between these scenarios, it is essential to probe the planet obliquity distribution
over a wide orbit separation range. 
One fundamental limitation for obtaining obliquity measurements with the RM effect is that 
this method can only be applied for transiting systems, i.e.\ systems with typical periods of a few days to tens of days.  

For wide-separation planets with astrometric orbits, the sky-projected obliquity angle can be derived from 
spectro-interferometric observations, such as provided by ESO's Very Large Telescope Interferometer (VLTI).
This method employs measuring wavelength-differential phases in photospheric absorption lines
in order to derive the photocenter shift associated with the stellar rotation.
Here, we present VLTI/GRAVITY high-spectral resolution data of the 
\object{$\beta$\,Pictoris} system, marking the second time that this technique is applied to a disk-hosting star 
(following the measurements that indicated good alignment of Fomalhaut's stellar equator 
with the orbit of its debris disk and dispersing dust cloud; \citealt{leb09,gas20})
and the first time that the obliquity angle is measured for a directly-imaged planetary system.

$\beta$\,Pic is a relatively young \citep[age $24\pm3$\,Myr,][]{bel15} A6V-type star 
at a distance of $19.45\pm0.05$\,pc \citep{van07} 
and with a mass of $1.85^{+0.03}_{-0.04}$\,M$_{\odot}$ \citep{wan16}.
\citep{def12} measured the stellar diameter to $0.736\pm0.019$\,mas 
and the projected rotation velocity is $v \sin i=130$\,km\,s$^{-1}$ \citep{roy07}. 
The system harbours a large-scale debris disk that exhibts a small-scale misalignment 
between the extended ``primary'' disk and a ``warped'' secondary disk in the inner 85\,au \citep{bur95,mou97,cur11,daw11}.

The planet \object{$\beta$\,Pic\,b} has been detected with adaptive optics imaging \citep{lag09a}
and has an orbital period of $\sim 20$\,yr with a semi-major axis of $10.6\pm0.5$\,au.
It is on a slightly inclined orbit with respect to the primary disk and 
might be responsible for triggering the warped secondary disk \citep{lag12}.
The orbit is seen with near-edge orientation ($i=89.04\pm0.03^{\circ}$, where 90$^{\circ}$ indicates edge-on viewing geometry) 
and the planet mass has been estimated to $12.8\pm2.2$\,M$_{\mathrm{Jup}}$ \citep{lag09b,fit09,wan16,sne18,gra20}.
\citet{lag19b} detected the radial velocity signal for a second $\sim 9$\,$M_{\mathrm{Jup}}$ planet,
$\beta$\,Pic\,c, on a $\sim 1,200$\,day orbit, corresponding to a semi-major axis of $\sim 2.7$\,au.

In Sect.~\ref{sec:observations} we describe our observational setup,
followed by a presentation of our modelling results (Sect.~\ref{sec:results}),
a discussion of the implications (Sect.~\ref{sec:discussion}) and  
our conclusions (Sect.~\ref{sec:conclusions}).

\section{Observations}
\label{sec:observations}

\begin{figure}[h]
  \centering
  \includegraphics[angle=0,scale=0.47]{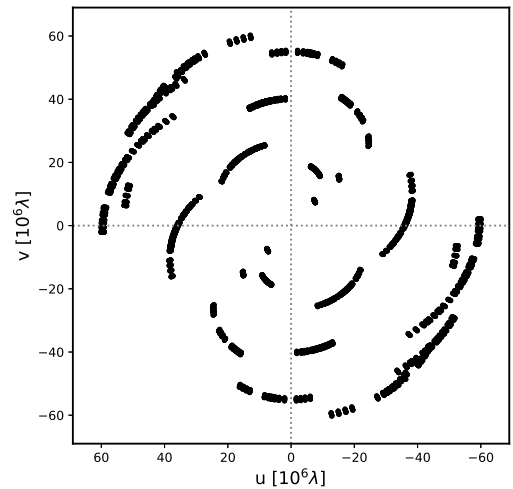} \\   
  \caption{ 
    uv-plane of VLTI/GRAVITY observations on $\beta$\,Pic.
    \label{fig:uvplane}
  }
\end{figure}

We observed $\beta$\,Pic with the GRAVITY instrument on seven nights 
between 2016 September 15 and 2016 October 20.
The observations combined the light from the four VLTI 1.8m auxiliary telescopes.  
The telescopes were located on stations 
A0-G1-J3-K0 (2016 September 15, 17, 18, 19),
A0-B2-J2-J3 (2016 October 16), and
A0-G1-J2-J3 (2016 October 18 and 20),
resulting in projected baseline lengths between 22.93 and 132.36\,m (Fig.~\ref{fig:uvplane}).

The interferograms cover the $K$-band with spectral resolution 
$\lambda/\Delta\lambda=4,000$ and were recorded with
detector integration times of 10\,s.
In total, we recorded 57 exposures of 300\,s on-source integration time on $\beta$\,Pic.
Interlayed with the observations on the science star,
we observed the calibrator star \object{HD\,159868}. 

Wavelength-differential visibilities and phases were extracted using the 
GRAVITY pipeline \citep[Release 1.2.4,][]{lap14}. 
In order to increase the signal-to-noise, we averaged the complex visibilities from
the 30 individual frames within an exposure.
The sky-projected baseline vectors keep changing within the 5\,min exposures due to Earth rotation
(at $\lesssim 0.2^{\circ}$/minute), which could potentially smear the measured astrometric signal.
However, based on the maximum measured astrometric displacement of $8\mu$as (Sect.~\ref{sec:results})
we estimate that the astrometric error induced by the averaging is less than $0.1\mu$as
and therefore negligible in our modeling.  
Furthermore, we note that this effect would only reduce the amplitude of the derived 
astrometric displacement without biasing the position angles of the derived astrometry vectors,
even if the averaging would be applied over a longer time interval.

\begin{figure}[h]
  \centering
  $\begin{array}{c}
    \includegraphics[angle=0,scale=0.5]{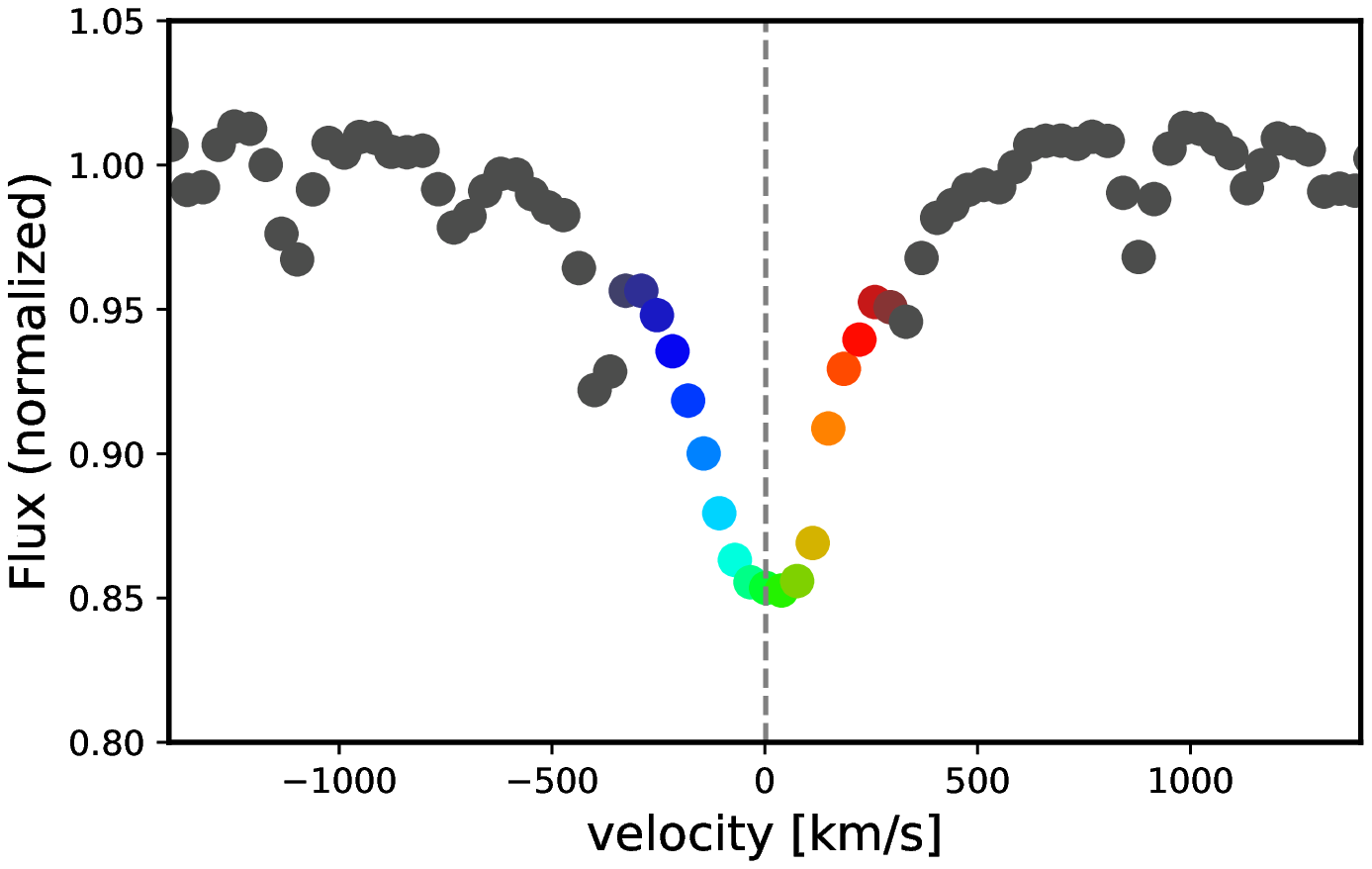} \\   
    \includegraphics[angle=0,scale=0.5]{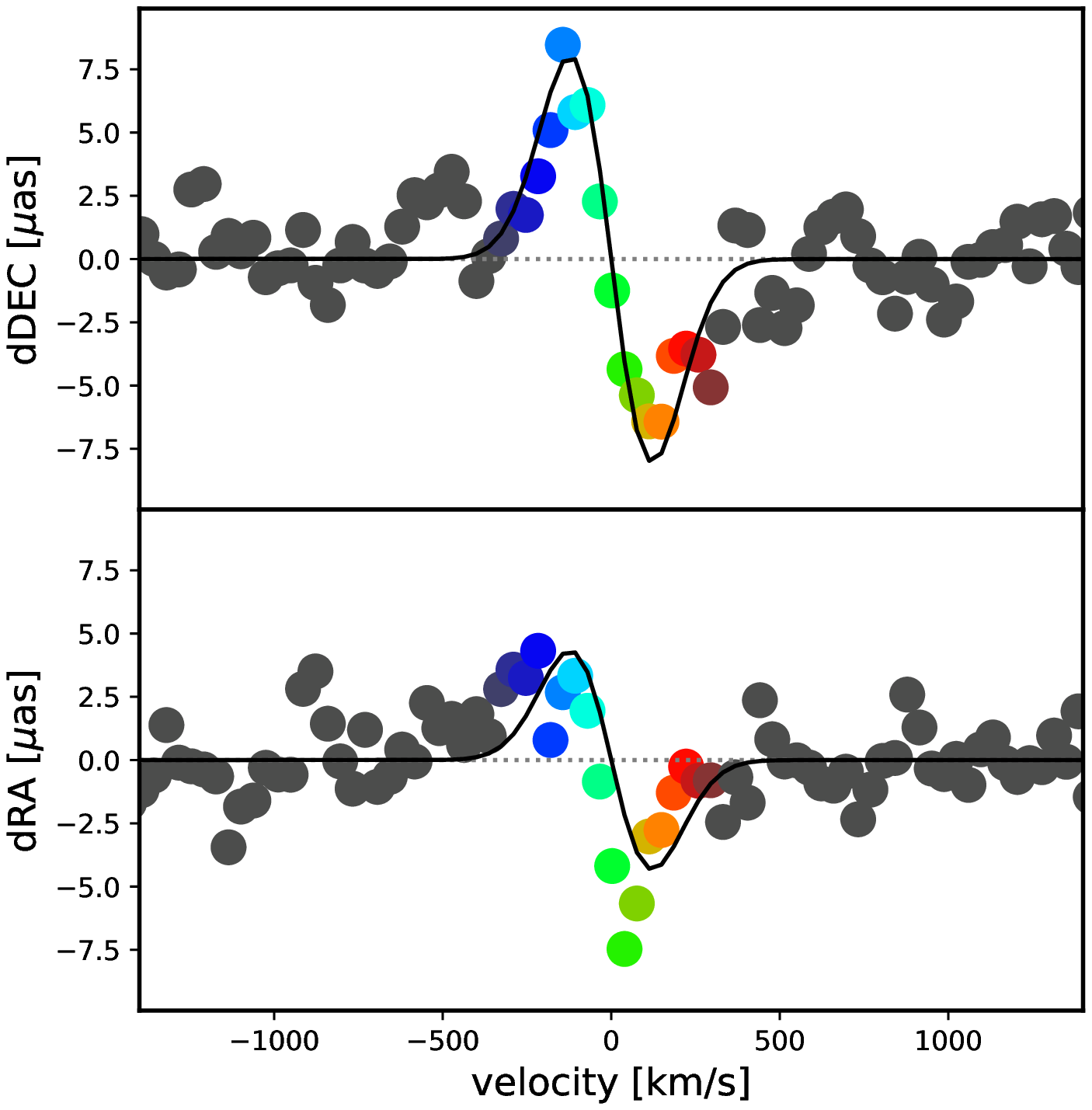}     
  \end{array}$
  \caption{
    Spectrum (top) and position-velocity diagrams along North-South and East-West direction (bottom).
    \label{fig:astrometry}
  }
\end{figure}

\begin{figure*}[h]
  \centering
    \includegraphics[angle=0,scale=0.9]{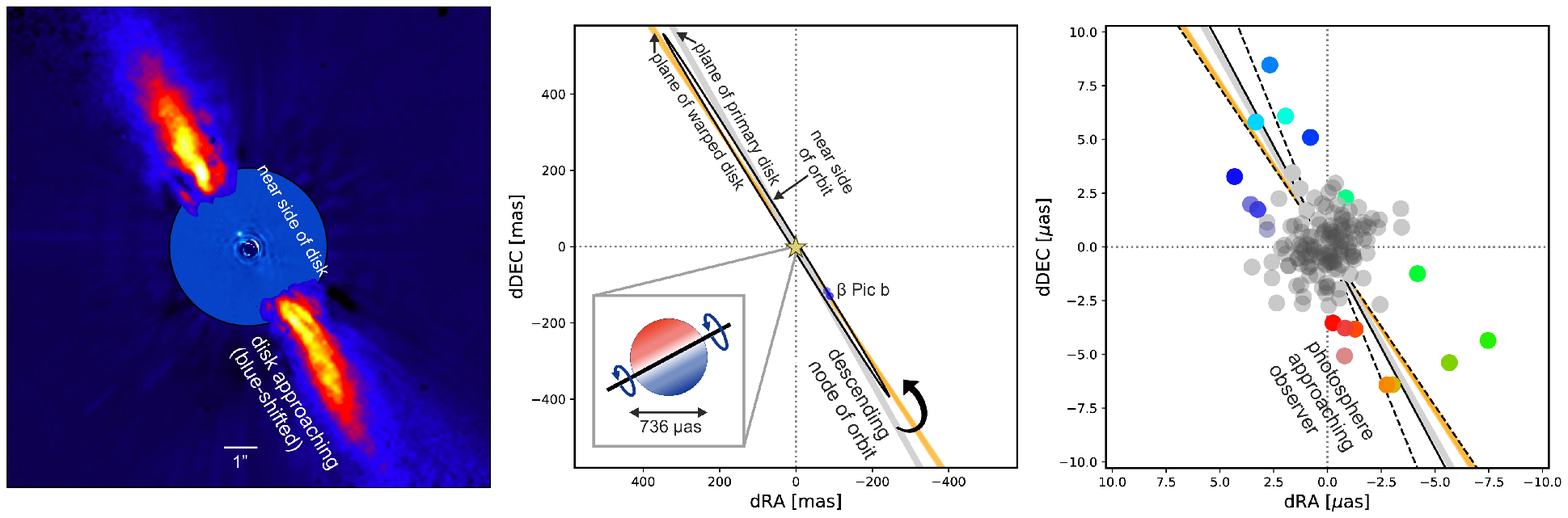}  
  \caption{
    {\it Left:} Scattered light image of $\beta$\,Pic's large-scale debris disk and of $\beta$\,Pic\,b \citep[credit: ESO,][]{lag09a},
    with the CO blue-shifted disk lobe and the side of the disk facing towards the observer indicated.
    {\it Middle:} The orbit of $\beta$\,Pic\,b \citep[orbital elements from][]{gra20}, with the plane of the primary disk
    and the plane of the warped disk marked as grey and orange line, respectively.  
    The position of $\beta$\,Pic\,b at the time of our observations has been marked with blue points.
    {\it Right:} Photocenter displacements measured in the Br$\gamma$ line (color points)
    and in the continuum (grey points).  
    The line channels are colored based on their wavelength, using the same color-coding as Fig.~\ref{fig:astrometry}.
    Given that the line is in {\it absorption}, the photocenters in the red-shifting wing of the line are displaced 
    towards the side of photosphere that is {\it approaching} the observer (and equivalent for the blue-shifted ling wing
    and the receeding side of the star).
    The solid black line gives the best-fit position angle for the 
    equatorial plane of the star, with the dashed black lines giving the 1$\sigma$ error intervals.
    In all panels, North is top and East is left.
    \label{fig:overview}
}
\end{figure*}

In individual exposures, we achieve a phase rms (root mean square) down to $0.39^{\circ}$ 
in the continuum channels near the Br$\gamma$ 2.1667\,$\mu$m line.
We fit the data from the 57 exposures simultaneously in order to reduce the noise further,
resulting in an overall differential astrometric accuracy of $1\mu$as
(as derived from the rms scatter in the continuum channels; grey points in Fig.~\ref{fig:overview}, right).
We reject two observations that exhibit phase rms $>0.8^{\circ}$ due to poor atmospheric conditions.
The spectrum of the pressure-broadened Br$\gamma$ line is shown in Fig.~\ref{fig:astrometry} (top).

\begin{figure*}[h]
  \centering
  \includegraphics[angle=0,scale=0.37]{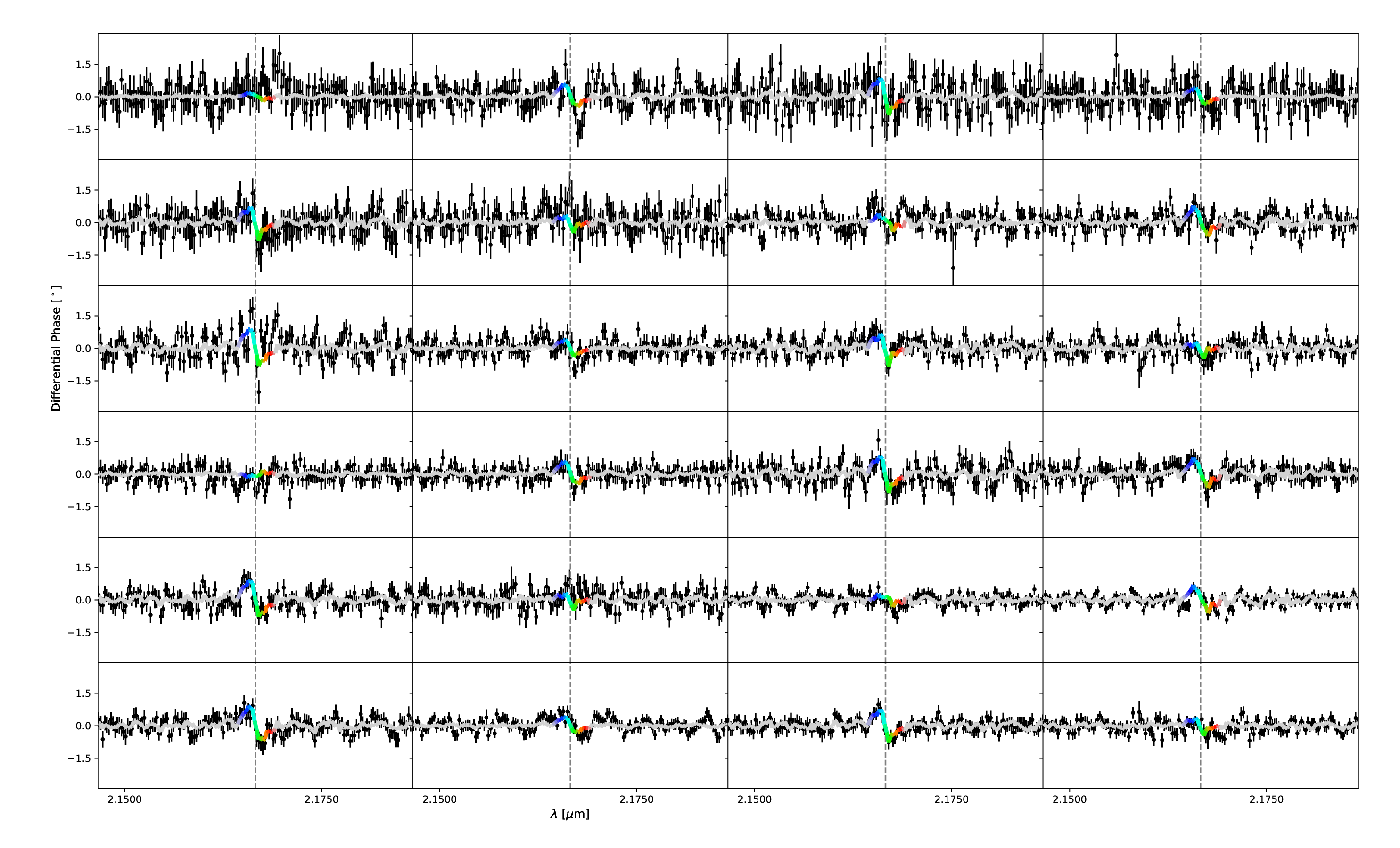}  
  \caption{ 
    Measured differential phases (points), overplotted with the model phases that correspond to our best-fit photocenter model (line).
    Each panel shows the measured phases for one of the 6 baselines, averaged over the full data recording sequence.
    \label{fig:DP}
  }
\end{figure*}

\section{Results}
\label{sec:results}

We infer the orientation of $\beta$\,Pic's stellar rotation axis from the spatial displacement 
in the photocenter between the blue- and red-shifted wing of the photospheric Br$\gamma$ absorption line.
The photocenter displacement in the line wings with respect to the continuum ``center of light'' is
just $\sim8\,\mu$as or $\sim$1/100th of the stellar diameter, but it can been measured with high significance 
from our wavelength-differential phases.

The wavelength-differential phase $\phi(\lambda)$ in a given spectral channel, $\lambda$,
and on a given baseline vector $\mathbf{B}$ at projected baseline coordinates $(u,v)$
is given by the equation 
\begin{equation}
  \phi(\lambda) = \frac{2\pi}{\lambda} \mathbf{B} \cdot \mathbf{x(\lambda)},   \label{eq:DP}
\end{equation}
where $\lambda$ is the wavelength of the considered spectral channel and $\mathbf{x(\lambda)}$ the 
photocenter displacement vector in the channel with respect to the continuum \citep{leb09}.

To maximise the signal-to-noise on the derived photocenter displacement vectors $\mathbf{x(\lambda)}$
for a given spectral channel $\lambda$, we arrange the $n$ measured differential phases in vector 
$\mathbf{\phi(\lambda)}$ and the corresponding baseline vectors in the $n\times 2$-dimensional matrix $\mathit{B}$.
Using the inverse matrix of $\mathit{B}$ we can rewrite eq.\,\ref{eq:DP} as:
\begin{equation}
  \mathbf{x(\lambda)} = \frac{\lambda}{2\pi} \mathit{B}^{-1} \mathbf{\phi(\lambda)^{\top}}
\end{equation}

Applying this equation to our data results in the photocenter vectors shown in Fig.~\ref{fig:overview}.
The right ascension and declination components of these vectors are also shown as position-velocity diagrams
in Fig.~\ref{fig:astrometry} (bottom).
The data show that the photocenter in the blue-shifted wing of the Br$\gamma$ line 
is displaced to the North-East (Fig.~\ref{fig:overview}, right).
As the line is in absorption, this indicates that the blue-shifted absorption is strongest on the
south-western side of the star, and thus the photocenter shifts to the North-East.
Accordingly, the blue-shifted (approaching) side of the rotating photosphere is towards the South-West.
The phases predicted by the model are overplotted on the measured differential phases in Fig.~\ref{fig:DP}.\\

In order to determine the position angle of the stellar rotation axis quantitatively,
we parameterize the wavelength-dependence of the photocenter displacement
between the blue- and red-shifted line wing with a S-shaped function $S(\lambda)=a(\lambda) \cdot \exp(-a(\lambda)^2)$,
where $a(\lambda)=2(\lambda-\lambda_{\mathrm{Br}\gamma})/w_{\mathrm{Br}\gamma}$, and 
$\lambda_{\mathrm{Br}\gamma}$ and $w_{\mathrm{Br}\gamma}$ are the central wavelength and width of the line.
This function provides a good approximation for the photocenter displacement 
due to rotation in a marginally resolved photosphere.
We fit this function to the astrometry vectors $\mathbf{x(\lambda)}$ in the complex plane:
\begin{equation}
  \mathbf{x(\lambda)} = c \cdot S(\lambda),
\end{equation}
where $c$ is a complex number.
The phase of $c$ gives the position angle of the spatial displacement between the blue- and red-shifted line wings, 
i.e.\ the stellar equator is oriented along $\Omega_\mathrm{s} = \tan^{-1}(\Re{c}/\Im{c})$.
We determine the uncertainties through bootstrapping.

For our data set on $\beta$\,Pic, we derive the sky-projected position angle of the equator to 
$\Omega_\mathrm{s}=29\pm4^{\circ}$, with the south-western side of the photosphere approaching 
the observer\footnote{We follow the convention to measure position angles East of North.}.
The bottom panel of Fig.~\ref{fig:astrometry} shows the model function overplotted on the measured astrometric signals.

There are some small residuals between model and data that might indicate a tiny (1-2\,$\mu$as) offset between the line photocenter 
and the continuum photocenter to the West (see red-shifted wing in the bottom panel of Fig.~\ref{fig:astrometry}).
Such an offset could indicate that the continuum photocenter is slightly shifted with respect to the center of the stellar photosphere, 
for instance due to stellar surface structures, diffuse disk emission, or a point source within the $\sim 0.1"$ field-of-view. 
However, the residuals have a low significance and data with higher signal-to-noise and higher angular resolution 
is needed to investigate this further.
Also, future differential phases measurements at higher spectral resolution could derive
an independent estimate for the inclination of the photosphere 
\citep[see model images of a inclined rotating photosphere in Fig.~8 of][]{kra12a} 
or detect subtle effects in the velocity field, such as related to differential rotation. 
However, we do not expect these effects to contribute to the residuals in our measurements.

\section{Discussion}
\label{sec:discussion}

Together with the inclination estimate of $i_{\mathrm{s}}=89.1^{\circ}$
derived from the pulsation frequency spectrum\footnote{The existing astroseismology 
measurements constrains the inclination value, but provide no information on the hemisphere
that faces towards Earth, resulting in a deneracy between $i_{\mathrm{s}}$ and $180-i_{\mathrm{s}}$.
Due to the near-edge viewing geometry, this degeneracy adds only a small uncertainty
that we take into account when computing mutual inclinations. } 
\citep{zwi19},
our measurement of $\Omega_\mathrm{s}$
defines the full spatial orientation of the stellar rotation axis,
enabling a comparison with the spatial orientation of the planetary orbit and of
the large-scale debris disk.

The primary disk is likely see nearly exactly edge-on, while the warped disk is tilted by $+4.0\pm0.6^{\circ}$ in position angle
with respect to the primary disk \citep{gol06,lag12} and inclined by $6\pm1^{\circ}$ with respect to the line of sight, with
the north-western part of the disk being tipped nearer to Earth \citep[][see Fig.~\ref{fig:overview}, left and middle]{ahm09}.
In order to compute the mutual inclination between the different planes,
we adopt the orbital elements of $\beta$\,Pic\,b published by \citet{gra20},
including an orbital inclination $i_{\mathrm{p}}=89.04\pm0.03^{\circ}$ and the 
longitude of the ascending node $\Omega_{\mathrm{p}}=31.88\pm0.05^{\circ}$.
For the primary disk plane we adopt $\Omega_{\mathrm{d,p}}=29.3^{+0.2}_{-0.3}$$^{\circ}$ \citep{lag12}
and $i_{\mathrm{d,p}}=90.0\pm 0.1$$^{\circ}$ \citep{ahm09}.
The warped secondary disk is oriented along $\Omega_{\mathrm{d,s}}=33.3\pm0.6^{\circ}$ \citep{lag12},  a slight tilt \citep[$i_{\mathrm{d,s}}=84.0\pm 1.0$$^{\circ}$,][]{ahm09}
that causes the north-western side of the disk to face towards the observer. 
This minimal tilt has been deduced from GPI scattered light imaging that show excess scattered light 
from the north-western side of the disk, likely due to dust forward scattering \citep{mil15}.

The direction of rotation of the disk has been measured by \citet{den14}.
They detected a molecular gas clump at a separation at $\sim85$\,au that might have been 
produced by the collision of icy bodies in $\beta$\,Pic's debris belt.  
The kinematics of the CO clump defines the rotation direction of the disk, 
with the approaching (blue-shifted) line wing to the South-West.
Accordingly, the rotation direction deduced from our observations matches the rotation direction 
for the large-scale disk.  Also, it matches the direction of rotation that has been deduced for $\beta$\,Pic\,b,
indicating that the planet is on a prograde orbit.

The sky-projected orientation of the different planes is shown in Fig.~\ref{fig:overview}.
The orbital plane of $\beta$\,Pic\,b is intermediate between the plane
of the primary disk (grey line in Fig.~\ref{fig:overview}) and of the warp disk (orange line),
as already noted by \citet{lag12}.  

Our measurements determine the sky-projected obliquity angle to $3\pm4^{\circ}$.
We compute the mutual inclination angle $\psi$ between the angular momentum vectors of
the stellar photosphere (s) and the planetary orbit (p) by adopting the equation from \citet{fek81}
\begin{equation}
\cos \psi = \cos i_\mathrm{s} \cos i_\mathrm{p} + \sin i_\mathrm{s} \sin i_\mathrm{p} \cos ( \Omega_\mathrm{p} - \Omega_\mathrm{s} ),
\label{eqn:mutualinclination}
\end{equation}
which yields $\psi_{\mathrm{p}} = 3\pm 5^{\circ}$ for the planetary orbit.
For the primary and secondary disk, the obliquity angles are 
$\psi_{\mathrm{d,p}} = 1\pm 4^{\circ}$ and $\psi_{\mathrm{d,s}} = 7\pm 8^{\circ}$, respectively.
This shows that the orbital planes both of the planet and of the primary debris disk
are well aligned with the stellar spin, suggesting that the planet formed in a 
coplanar disk whose angular momentum vector was well-aligned with the stellar rotation axis.
Any migration to the planet's current location cannot have occured via mechanisms 
that would have inflated its mutual inclination.\\

Our observation provides a first glimpse on the spin-orbit alignment distribution for wide-separation
planets, adding new constraints that will help to test hypotheses that have been put forward 
to explain the origin of the obliquity in short-period planet systems.
For the case of $\beta$\,Pic\,b, we can rule out scenarios that induce 
stellar obliquities as part of the star formation process or during the disk evolution.
The mechanisms that have been proposed to produce such primordial misalignments 
include planet formation either in a non-coplanar (i.e.\ warped) disk or in a disk whose 
angular momentum vector has been shifted with respect to the rotation axis of the star, 
for instance due to magnetic \citep{lai11} or hydrodynamical effects \citep{rog13}, or
due to turbulence in the cloud that formed the star initially \citep{bat10,thi11,fie15}.
A key prediction of these primodial formation scenarios is that they can induce
obliquities not only for short-separation, but also wide-separation planets,
such as $\beta$\,Pic\,b.

\section{Conclusions}
\label{sec:conclusions} 

We measured the sky-projected obliquity angle for the planet-host star $\beta$\,Pictoris to $3\pm4^{\circ}$.
Incorporating the inclination constraints derived with astroseismology
we constrain the 3-dimensional orientation of the stellar rotation axis and determine the 
mutual inclination angle between the angular momentum vector of the star and of 
$\beta$\,Pic\,b to $3\pm5^{\circ}$, indicating that the stellar spin axis is 
well aligned with the planetary orbit, as well as with the primary disk.

Our finding of spin-orbit alignment for $\beta$\,Pic\,b suggests that this planet 
formed in a coplanar disks without primordial misalignments.
This is in contrast to theories that describe the occurence of obliquities as a 
natural by-product of the star formation process, for instance through turbulent motions 
in the star-forming cloud \citep{fie15} or magnetic \citep{lai11} and fluid-dynamical effects \citep{rog13} during disk formation.
In case our finding of a well-aligned system is representative for wide-separation planets, 
it would suggest that the population of Hot Jupiters on oblique orbits (found in RM survey 
at orbit separations between $\sim 0.02$ and 0.3\,au) are likely transferred to
oblique orbits through dynamical processes post-formation.
Possible mechanisms include planet-planet scattering, stellar flybys,
or the Kozai-Lidov mechanism, where a wide companion orbiting a close binary 
on a highly inclined orbit can induce oscillations in inclination/eccentricity of the close pair \citep[e.g.][]{fab07}.

Observations on a larger sample of planet-hosting stars will be critical to test this hypothesis
and to extend the obliquity distribution presently covered by RM-based surveys ($\lesssim 0.3$\,au) out to planets with orbital separations at tens or hundreds of au.
For such wide-separation systems, infrared interferometry at high spectral dispersion
provides the only technique to measure the sky-projected obliquity angle.
At present, spin-orbit alignment measurements with VLTI are limited to nearby stars
with large apparent diameters and to pressure-broadened lines of relatively fast-rotating stars,
which strongly limits the numbers of stars that are accessible with this technique.
A dedicated high-spectral resolution ($R=25,000$), short-wavelength instrument
operating in the J-band (1-1.4$\mu$m) and optimised for precision phase measurements,
such as the proposed VLTI visitor instrument BIFROST \citep{kra19},
will be able to mitigate these limitations and enable spin-orbit measurements for 
hundreds of systems in the planet samples that are expected from 
direct-imaging facilities (JWST and ELTs) and the GAIA astrometry mission.
Modeling the obliquity distribution that will be provided by such next-generation instruments
will allow testing of the theories that have been put forward to explain the origin of 
planet obliquity on a statistically significant sample, offering direct insights on the planet formation 
process and the dynamical evolution that shapes the architecture of planetary systems.

\acknowledgments

We thank Gilles Duvert for helpful input on GRAVITY data reduction.
We acknowledge support from an ERC Starting grant (Grant Agreement \#639889).
J.D.M.\ acknowledges funding from National Science Foundation grant NSF-AST1506540 and NASA grant NNX16AD43G. 
G.M.K.\ is supported by the Royal Society as a Royal Society University Research fellow.
Based on observations made with ESO telescopes at Paranal Observatory under program IDs 60.A-9165(A) and 098.C-0702(A).    
Some of the presented data has been recorded as part of a GRAVITY Science Verification program
and we thank the SV team, which is composed of ESO employees and GRAVITY consortium members,
for carrying out the observations (https://www.eso.org/sci/activities/vltsv/gravitysv.html).

{\it Facilities:} \facility{VLTI}


\begin{thebibliography}{44}
\expandafter\ifx\csname natexlab\endcsname\relax\def\natexlab#1{#1}\fi

\bibitem[{{Ahmic} {et~al.}(2009){Ahmic}, {Croll}, \& {Artymowicz}}]{ahm09}
{Ahmic}, M., {Croll}, B., \& {Artymowicz}, P. 2009, \apj, 705, 529

\bibitem[{{Bate} {et~al.}(2010){Bate}, {Lodato}, \& {Pringle}}]{bat10}
{Bate}, M.~R., {Lodato}, G., \& {Pringle}, J.~E. 2010, \mnras, 401, 1505

\bibitem[{{Beck} \& {Giles}(2005)}]{bec05}
{Beck}, J.~G., \& {Giles}, P. 2005, \apjl, 621, L153

\bibitem[{{Bell} {et~al.}(2015){Bell}, {Mamajek}, \& {Naylor}}]{bel15}
{Bell}, C. P.~M., {Mamajek}, E.~E., \& {Naylor}, T. 2015, \mnras, 454, 593

\bibitem[{{Burrows} {et~al.}(1995){Burrows}, {Krist}, {Stapelfeldt}, \& {WFPC2
  Investigation Definition Team}}]{bur95}
{Burrows}, C.~J., {Krist}, J.~E., {Stapelfeldt}, K.~R., \& {WFPC2 Investigation
  Definition Team}. 1995, in American Astronomical Society Meeting Abstracts,
  Vol. 187, American Astronomical Society Meeting Abstracts, 32.05

\bibitem[{{Chatterjee} {et~al.}(2008){Chatterjee}, {Ford}, {Matsumura}, \&
  {Rasio}}]{cha08}
{Chatterjee}, S., {Ford}, E.~B., {Matsumura}, S., \& {Rasio}, F.~A. 2008, \apj,
  686, 580

\bibitem[{{Currie} {et~al.}(2011){Currie}, {Thalmann}, {Matsumura},
  {Madhusudhan}, {Burrows}, \& {Kuchner}}]{cur11}
{Currie}, T., {Thalmann}, C., {Matsumura}, S., {Madhusudhan}, N., {Burrows},
  A., \& {Kuchner}, M. 2011, \apjl, 736, L33

\bibitem[{{Dawson} {et~al.}(2011){Dawson}, {Murray-Clay}, \&
  {Fabrycky}}]{daw11}
{Dawson}, R.~I., {Murray-Clay}, R.~A., \& {Fabrycky}, D.~C. 2011, \apjl, 743,
  L17

\bibitem[{{Defr{\`e}re} {et~al.}(2012){Defr{\`e}re}, {Lebreton}, {Le Bouquin},
  {Lagrange}, {Absil}, {Augereau}, {Berger}, {di Folco}, {Ertel}, {Kluska},
  {Montagnier}, {Millan-Gabet}, {Traub}, \& {Zins}}]{def12}
{Defr{\`e}re}, D., {et~al.} 2012, \aap, 546, L9

\bibitem[{{Dent} {et~al.}(2014){Dent}, {Wyatt}, {Roberge}, {Augereau},
  {Casassus}, {Corder}, {Greaves}, {de Gregorio-Monsalvo}, {Hales}, {Jackson},
  {Hughes}, {Lagrange}, {Matthews}, \& {Wilner}}]{den14}
{Dent}, W.~R.~F., {et~al.} 2014, Science, 343, 1490

\bibitem[{{Fabrycky} \& {Tremaine}(2007)}]{fab07}
{Fabrycky}, D., \& {Tremaine}, S. 2007, \apj, 669, 1298

\bibitem[{{Fekel}(1981)}]{fek81}
{Fekel}, F.~C., J. 1981, \apj, 246, 879

\bibitem[{{Fielding} {et~al.}(2015){Fielding}, {McKee}, {Socrates},
  {Cunningham}, \& {Klein}}]{fie15}
{Fielding}, D.~B., {McKee}, C.~F., {Socrates}, A., {Cunningham}, A.~J., \&
  {Klein}, R.~I. 2015, \mnras, 450, 3306

\bibitem[{{Fitzgerald} {et~al.}(2009){Fitzgerald}, {Kalas}, \&
  {Graham}}]{fit09}
{Fitzgerald}, M.~P., {Kalas}, P.~G., \& {Graham}, J.~R. 2009, \apjl, 706, L41

\bibitem[{{Gaspar} \& {Rieke}(2020)}]{gas20}
{Gaspar}, A., \& {Rieke}, G. 2020, Proceedings of the National Academy of
  Science, 117, 9712

\bibitem[{{Golimowski} {et~al.}(2006){Golimowski}, {Ardila}, {Krist},
  {Clampin}, {Ford}, {Illingworth}, {Bartko}, {Ben{\'\i}tez}, {Blakeslee},
  {Bouwens}, {Bradley}, {Broadhurst}, {Brown}, {Burrows}, {Cheng}, {Cross},
  {Demarco}, {Feldman}, {Franx}, {Goto}, {Gronwall}, {Hartig}, {Holden},
  {Homeier}, {Infante}, {Jee}, {Kimble}, {Lesser}, {Martel}, {Mei},
  {Menanteau}, {Meurer}, {Miley}, {Motta}, {Postman}, {Rosati}, {Sirianni},
  {Sparks}, {Tran}, {Tsvetanov}, {White}, {Zheng}, \& {Zirm}}]{gol06}
{Golimowski}, D.~A., {et~al.} 2006, \aj, 131, 3109

\bibitem[{{Gravity Collaboration} {et~al.}(2020){Gravity Collaboration},
  {Nowak}, {Lacour}, {Molli{\`e}re}, {Wang}, {Charnay}, {van Dishoeck},
  {Abuter}, {Amorim}, {Berger}, {Beust}, {Bonnefoy}, {Bonnet}, {Brandner},
  {Buron}, {Cantalloube}, {Collin}, {Chapron}, {Cl{\'e}net}, {Coud{\'e} Du
  Foresto}, {de Zeeuw}, {Dembet}, {Dexter}, {Duvert}, {Eckart}, {Eisenhauer},
  {F{\"o}rster Schreiber}, {F{\'e}dou}, {Garcia Lopez}, {Gao}, {Gendron},
  {Genzel}, {Gillessen}, {Hau{\ss}mann}, {Henning}, {Hippler}, {Hubert},
  {Jocou}, {Kervella}, {Lagrange}, {Lapeyr{\`e}re}, {Le Bouquin}, {L{\'e}na},
  {Maire}, {Ott}, {Paumard}, {Paladini}, {Perraut}, {Perrin}, {Pueyo}, {Pfuhl},
  {Rabien}, {Rau}, {Rodr{\'\i}guez-Coira}, {Rousset}, {Scheithauer},
  {Shangguan}, {Straub}, {Straubmeier}, {Sturm}, {Tacconi}, {Vincent},
  {Widmann}, {Wieprecht}, {Wiezorrek}, {Woillez}, {Yazici}, \&
  {Ziegler}}]{gra20}
{Gravity Collaboration} {et~al.} 2020, \aap, 633, A110

\bibitem[{{H{\'e}brard} {et~al.}(2011){H{\'e}brard}, {Ehrenreich}, {Bouchy},
  {Delfosse}, {Moutou}, {Arnold}, {Boisse}, {Bonfils}, {D{\'\i}az},
  {Eggenberger}, {Forveille}, {Lagrange}, {Lovis}, {Pepe}, {Perrier}, {Queloz},
  {Santerne}, {Santos}, {S{\'e}gransan}, {Udry}, \& {Vidal-Madjar}}]{heb11}
{H{\'e}brard}, G., {et~al.} 2011, \aap, 527, L11

\bibitem[{{Kozai}(1962)}]{koz62}
{Kozai}, Y. 1962, \aj, 67, 591

\bibitem[{{Kraus}(2019)}]{kra19}
{Kraus}, S. 2019, in The Very Large Telescope in 2030, 36

\bibitem[{{Kraus} {et~al.}(2012){Kraus}, {Monnier}, {Che}, {Schaefer},
  {Touhami}, {Gies}, {Aufdenberg}, {Baron}, {Thureau}, {ten Brummelaar},
  {McAlister}, {Turner}, {Sturmann}, \& {Sturmann}}]{kra12a}
{Kraus}, S., {et~al.} 2012, \apj, 744, 19

\bibitem[{{Kraus} {et~al.}(2020){Kraus}, {Kreplin}, {Young}, {Bate}, {Monnier},
  {Harries}, {Avenhaus}, {Kluska}, {Laws}, {Rich}, {Willson}, {Aarnio},
  {Adams}, {Andrews}, {Anugu}, {Bae}, {ten Brummelaar}, {Calvet}, {Cur{\'e}},
  {Davies}, {Ennis}, {Espaillat}, {Gardner}, {Hartmann}, {Hinkley}, {Labdon},
  {Lanthermann}, {LeBouquin}, {Schaefer}, {Setterholm}, {Wilner}, \&
  {Zhu}}]{kra20}
---. 2020, arXiv e-prints, arXiv:2004.01204

\bibitem[{{Lagrange} {et~al.}(2009{\natexlab{a}}){Lagrange}, {Gratadour},
  {Chauvin}, {Fusco}, {Ehrenreich}, {Mouillet}, {Rousset}, {Rouan}, {Allard},
  {Gendron}, {Charton}, {Mugnier}, {Rabou}, {Montri}, \& {Lacombe}}]{lag09a}
{Lagrange}, A.~M., {et~al.} 2009{\natexlab{a}}, \aap, 493, L21

\bibitem[{{Lagrange} {et~al.}(2009{\natexlab{b}}){Lagrange}, {Kasper},
  {Boccaletti}, {Chauvin}, {Gratadour}, {Fusco}, {Ehrenreich}, {Apai},
  {Mouillet}, \& {Rouan}}]{lag09b}
---. 2009{\natexlab{b}}, \aap, 506, 927

\bibitem[{{Lagrange} {et~al.}(2012){Lagrange}, {Boccaletti}, {Milli},
  {Chauvin}, {Bonnefoy}, {Mouillet}, {Augereau}, {Girard}, {Lacour}, \&
  {Apai}}]{lag12}
---. 2012, \aap, 542, A40

\bibitem[{{Lagrange} {et~al.}(2019){Lagrange}, {Meunier}, {Rubini}, {Keppler},
  {Galland}, {Chapellier}, {Michel}, {Balona}, {Beust}, {Guillot}, {Grandjean},
  {Borgniet}, {M{\'e}karnia}, {Wilson}, {Kiefer}, {Bonnefoy}, {Lillo-Box},
  {Pantoja}, {Jones}, {Iglesias}, {Rodet}, {Diaz}, {Zapata}, {Abe}, \&
  {Schmider}}]{lag19b}
---. 2019, Nature Astronomy, 3, 1135

\bibitem[{{Lai} {et~al.}(2011){Lai}, {Foucart}, \& {Lin}}]{lai11}
{Lai}, D., {Foucart}, F., \& {Lin}, D.~N.~C. 2011, \mnras, 412, 2790

\bibitem[{{Lapeyrere} {et~al.}(2014){Lapeyrere}, {Kervella}, {Lacour},
  {Azouaoui}, {Garcia-Dabo}, {Perrin}, {Eisenhauer}, {Perraut}, {Straubmeier},
  {Amorim}, \& {Brandner}}]{lap14}
{Lapeyrere}, V., {et~al.} 2014, in \procspie, Vol. 9146, Optical and Infrared
  Interferometry IV, 91462D

\bibitem[{{Le Bouquin} {et~al.}(2009){Le Bouquin}, {Absil}, {Benisty}, {Massi},
  {M{\'e}rand}, \& {Stefl}}]{leb09}
{Le Bouquin}, J.-B., {Absil}, O., {Benisty}, M., {Massi}, F., {M{\'e}rand}, A.,
  \& {Stefl}, S. 2009, \aap, 498, L41

\bibitem[{{Millar-Blanchaer} {et~al.}(2015){Millar-Blanchaer}, {Graham},
  {Pueyo}, {Kalas}, {Dawson}, {Wang}, {Perrin}, {moon}, {Macintosh}, {Ammons},
  {Barman}, {Cardwell}, {Chen}, {Chiang}, {Chilcote}, {Cotten}, {De Rosa},
  {Draper}, {Dunn}, {Duch{\^e}ne}, {Esposito}, {Fitzgerald}, {Follette},
  {Goodsell}, {Greenbaum}, {Hartung}, {Hibon}, {Hinkley}, {Ingraham},
  {Jensen-Clem}, {Konopacky}, {Larkin}, {Long}, {Maire}, {Marchis}, {Marley},
  {Marois}, {Morzinski}, {Nielsen}, {Palmer}, {Oppenheimer}, {Poyneer},
  {Rajan}, {Rantakyr{\"o}}, {Ruffio}, {Sadakuni}, {Saddlemyer}, {Schneider},
  {Sivaramakrishnan}, {Soummer}, {Thomas}, {Vasisht}, {Vega}, {Wallace},
  {Ward-Duong}, {Wiktorowicz}, \& {Wolff}}]{mil15}
{Millar-Blanchaer}, M.~A., {et~al.} 2015, \apj, 811, 18

\bibitem[{{Mouillet} {et~al.}(1997){Mouillet}, {Larwood}, {Papaloizou}, \&
  {Lagrange}}]{mou97}
{Mouillet}, D., {Larwood}, J.~D., {Papaloizou}, J.~C.~B., \& {Lagrange}, A.~M.
  1997, \mnras, 292, 896

\bibitem[{{Mu{\~n}oz} \& {Perets}(2018)}]{mun18}
{Mu{\~n}oz}, D.~J., \& {Perets}, H.~B. 2018, \aj, 156, 253

\bibitem[{{Nixon}(2012)}]{nix12}
{Nixon}, C.~J. 2012, \mnras, 423, 2597

\bibitem[{{Queloz} {et~al.}(2000){Queloz}, {Eggenberger}, {Mayor}, {Perrier},
  {Beuzit}, {Naef}, {Sivan}, \& {Udry}}]{que00}
{Queloz}, D., {Eggenberger}, A., {Mayor}, M., {Perrier}, C., {Beuzit}, J.~L.,
  {Naef}, D., {Sivan}, J.~P., \& {Udry}, S. 2000, \aap, 359, L13

\bibitem[{{Rogers} \& {Lin}(2013)}]{rog13}
{Rogers}, T.~M., \& {Lin}, D.~N.~C. 2013, \apjl, 769, L10

\bibitem[{{Royer} {et~al.}(2007){Royer}, {Zorec}, \& {G{\'o}mez}}]{roy07}
{Royer}, F., {Zorec}, J., \& {G{\'o}mez}, A.~E. 2007, \aap, 463, 671

\bibitem[{{Snellen} \& {Brown}(2018)}]{sne18}
{Snellen}, I.~A.~G., \& {Brown}, A.~G.~A. 2018, Nature Astronomy, 2, 883

\bibitem[{{Thies} {et~al.}(2011){Thies}, {Kroupa}, {Goodwin}, {Stamatellos}, \&
  {Whitworth}}]{thi11}
{Thies}, I., {Kroupa}, P., {Goodwin}, S.~P., {Stamatellos}, D., \& {Whitworth},
  A.~P. 2011, \mnras, 417, 1817

\bibitem[{{Valsecchi} \& {Rasio}(2014)}]{val14}
{Valsecchi}, F., \& {Rasio}, F.~A. 2014, \apj, 786, 102

\bibitem[{{van Leeuwen}(2007)}]{van07}
{van Leeuwen}, F. 2007, \aap, 474, 653

\bibitem[{{Wang} {et~al.}(2016){Wang}, {Graham}, {Pueyo}, {Kalas},
  {Millar-Blanchaer}, {Ruffio}, {De Rosa}, {Ammons}, {Arriaga}, {Bailey},
  {Barman}, {Bulger}, {Burrows}, {Cardwell}, {Chen}, {Chilcote}, {Cotten},
  {Fitzgerald}, {Follette}, {Doyon}, {Duch{\^e}ne}, {Greenbaum}, {Hibon},
  {Hung}, {Ingraham}, {Konopacky}, {Larkin}, {Macintosh}, {Maire}, {Marchis},
  {Marley}, {Marois}, {Metchev}, {Nielsen}, {Oppenheimer}, {Palmer}, {Patel},
  {Patience}, {Perrin}, {Poyneer}, {Rajan}, {Rameau}, {Rantakyr{\"o}},
  {Savransky}, {Sivaramakrishnan}, {Song}, {Soummer}, {Thomas}, {Vasisht},
  {Vega}, {Wallace}, {Ward-Duong}, {Wiktorowicz}, \& {Wolff}}]{wan16}
{Wang}, J.~J., {et~al.} 2016, \aj, 152, 97

\bibitem[{{Winn} {et~al.}(2010){Winn}, {Fabrycky}, {Albrecht}, \&
  {Johnson}}]{win10}
{Winn}, J.~N., {Fabrycky}, D., {Albrecht}, S., \& {Johnson}, J.~A. 2010, \apjl,
  718, L145

\bibitem[{{Wright} {et~al.}(2011){Wright}, {Chen{\'e}}, {De Cat}, {Marois},
  {Mathias}, {Macintosh}, {Isaacs}, {Lehmann}, \& {Hartmann}}]{wri11}
{Wright}, D.~J., {et~al.} 2011, \apjl, 728, L20

\bibitem[{{Zwintz} {et~al.}(2019){Zwintz}, {Reese}, {Neiner}, {Pigulski},
  {Kuschnig}, {M{\"u}llner}, {Zieba}, {Abe}, {Guillot}, {Handler}, {Kenworthy},
  {Stuik}, {Moffat}, {Popowicz}, {Rucinski}, {Wade}, {Weiss}, {Bailey},
  {Crawford}, {Ireland}, {Lomberg}, {Mamajek}, {Mellon}, \& {Talens}}]{zwi19}
{Zwintz}, K., {et~al.} 2019, \aap, 627, A28

\end{thebibliography}

\end{document}